\documentclass[sigconf,nonacm]{acmart}
\definecolor{MAGENTA}{rgb}{1.0, 0.0, 1.0} 
\definecolor{darkgreen}{rgb}{0.0, 0.7, 0.2} 
\definecolor{darkyellow}{rgb}{0.8, 0.7, 0.3} 
\usepackage{textcomp}%
\usepackage{booktabs}%
\usepackage{amsmath}%
\usepackage{rotating}%
\usepackage{array} 
\usepackage[utf8]{inputenc}
\usepackage{tabu}%
\usepackage{multirow}%
\usepackage{graphicx}
\usepackage{fancyvrb}
\usepackage{tcolorbox}
\usepackage{lipsum}

\newcommand{\llama}[0]{LLaMA}

\newcommand{\llmjudge}[0]{DL 2023}
\newcommand{\dltwenty}[0]{DL 2020}
\newcommand{\dlnineteen}[0]{DL 2019}
\newcommand{\llamasmall}[0]{\llama{}-3-8B}
\newcommand{\llamabig}[0]{\llama{}-3.3-70B}
\newcommand{\tfive}[0]{FLAN-T5-large}
\newcommand{\deepseek}[0]{DeepSeek V3}
\newcommand{\gpto}[0]{GPT-4o}

\usepackage{placeins}

\thanks{This paper has been accepted for publication at SIGIR 2025. This is a preprint version.}

\begin{document}

\title{Does UMBRELA Work on Other LLMs?}

\begin{abstract}
We reproduce the UMBRELA LLM Judge evaluation framework across a range of large language models (LLMs) to assess its generalizability beyond the original study. Our investigation evaluates how LLM choice affects relevance assessment accuracy, focusing on leaderboard rank correlation and per-label agreement metrics. Results demonstrate that UMBRELA with \deepseek{} obtains very comparable performance to \gpto{} (used in original work). For \llamabig{} we obtain slightly lower performance, which further degrades with smaller LLMs.%
\footnote{
\textbf{Appendix: \url{https://github.com/TREMA-UNH/appendix-umbrella-other-llm}}}
\end{abstract}

\author{Naghmeh Farzi}
\affiliation{%
    \institution{University of New Hampshire}
    \city{Durham}
    \country{USA}
}
\email{Naghmeh.Farzi@unh.edu}
  
\author{Laura Dietz}
\affiliation{%
    \institution{University of New Hampshire}
    \city{Durham}
    \country{USA}
}
\email{dietz@cs.unh.edu}


\maketitle

\section{Introduction}

Relevance labels play a critical role in training and evaluating retrieval systems. Traditionally, these labels have been assigned by human annotators, who assess the relevance of documents or passages to a query based on predefined and provided guidelines. While human annotations have been considered the gold standard, they are resource-intensive and time-consuming. As large language models (LLMs) such as GPT-4o \cite{gpt4o2024} gain prominence, researchers have explored using prompting techniques to predict relevance labels. In this approach, generally a query and passage are input together to an LLM, which generates a response that is parsed and mapped to a relevance label, aiding human annotation with a scalable solution.

Recent research using this technique has proposed methods of varying complexity, ranging from direct prompts for binary relevance assessment \cite{Sun2023IsCG,Faggioli2023PerspectivesOL} to more refined methods where LLMs predict numerical relevance labels \cite{Thomas2023LargeLM,upadhyay2024umbrelaumbrelaopensourcereproduction}. Among these, the UMBRELA framework \cite{upadhyay2024umbrelaumbrelaopensourcereproduction} stands out for its ease of use and effectiveness. UMBRELA employs a zero-shot prompting approach to produce graded relevance judgments. This framework has been widely adopted due to its high correlation with human annotations on TREC Deep Learning datasets. While it is technically compatible with any LLM, it remains an open question whether its effectiveness holds when applied with different models.

\paragraph{System to be reproduced: UMBRELA}

The UMBRELA framework (a recursive acronym for \textbf{UM}brela is the \textbf{B}ing \textbf{REL}evance \textbf{A}ssessor) \cite{upadhyay2024umbrelaumbrelaopensourcereproduction} is an automated relevance assessment system designed to evaluate search results using large language models (LLMs). At its core, UMBRELA employs a zero-shot DNA (Descriptive, Narrative, and Aspects) prompting \cite{Thomas2023LargeLM} approach (as detailed in Figure~\ref{fig:umbrela_prompt}) to generate graded relevance labels. By leveraging an explicitly structured prompt, the system directs the LLM to assess search relevance based on two primary aspects: (1) the alignment between the query’s intent and the passage content, and (2) the trustworthiness of the passage. This approach leverages \gpto{} to provide a scalable, cost-effective, and reproducible complement to human relevance assessment.

Experimental results validate UMBRELA's effectiveness, demonstrating high correlations with human judgments on TREC Deep Learning datasets (2019–2023) \cite{dl19,dl20,rahmani_llmjudge_2024}. With Kendall’s $\tau$ measurements exceeding 0.8728 across all datasets, UMBRELA exhibits agreement levels akin to human expert assessors. Due to its reliability and performance, UMBRELA has been adopted as an official component of the TREC Retrieval-Augmented Generation (RAG) track in 2024, further underscoring its value in modern information retrieval evaluations.  

Because UMBRELA can be used with any LLM, IR research relying on its judgments remains reproducible, even if \gpto{} becomes unavailable, by switching to models with publicly available weights. Additionally, UMBRELA’s status as a public domain implementation of DNA prompting approach proposed by Thomas et al. \cite{Thomas2023LargeLM}, designed to work across LLMs, enhances its accessibility and utility for the IR community. 

In this work, we apply UMBRELA's prompt to a range of LLM families and examine whether other LLMs can replace \gpto{} within the framework without performance degradation. While much of the existing research has focused on prompt design, the impact of the underlying LLM is often underexplored. In this work, we explore the connection between the prompt and the LLM family.

\paragraph{Contributions}

We evaluate UMBRELA's reproducibility across various LLM families, addressing the following research questions:

\begin{itemize}
    \item RQ1: Can we replace the LLM in UMBRELA without performance degradation?  
    \item RQ2: To what extent do leaderboard rank correlation and per-label agreement metrics differ in their sensitivity to LLM scale?  
    \item RQ3: How consistent are UMBRELA’s performance patterns across different relevance assessment datasets, and does this consistency vary with LLM scale?  
    \item RQ4: To what extent does investing in larger language models provide meaningful improvements in relevance assessment quality beyond leaderboard rank correlation?  
    \item RQ5: How much does the word choice of the prompt matter in comparison to the choice of LLM?  
\end{itemize}

\section{Methodology}
\subsection{Work to be Reproduced: UMBRELA's Approach}

UMBRELA is an LLM-based framework designed to generate graded relevance assessments in information retrieval. It operates through a single, structured prompt, given in Figure \ref{fig:umbrela_prompt}, which is devised to guide the model in assigning a relevance label to a query-passage pair without requiring explicit examples, few-shot learning, or external annotations.

At the core of UMBRELA is a zero-shot prompting strategy that instructs the model to evaluate the passage's relevance along a predefined four-point scale. The prompt defines relevance categories from $0$ (completely unrelated) to $3$ (fully relevant), ensuring that each score corresponds to an unambiguous interpretation of relevance. To further reinforce correct label assignment, the prompt explicitly differentiates between borderline cases, instructing the model to use score $1$ when the passage is related but does not answer the query, score $2$ when the passage contains relevant information but is diluted by extraneous content, and score $3$ when the passage is entirely dedicated to answering the query.

A critical aspect of UMBRELA's prompt design is that it implicitly encourages the model to assess relevance from multiple perspectives. Without enforcing a rigid step-by-step reasoning process, it introduces considerations such as query intent alignment. This directs the model to measure how well the passage matches the search intent. It also includes trustworthiness, which prompts the model to factor in the reliability of the passage as a source. These aspects are not explicitly separated into distinct reasoning steps but are rather only guided by the prompt, allowing the LLM to integrate them naturally while producing a final judgment.

A key design choice in UMBRELA's prompt is the requirement of a strict output format, ensuring consistency and machine-readability. The model is instructed to output only a single integer score in the form ``\texttt{\#\#final score: [score]}'' without explanation or additional text. This design choice supports efficient and reliable relevance assessment across large-scale search evaluations.

By combining the state-of-the-art \gpto{} LLM with a carefully optimized prompt, UMBRELA provides a scalable, reproducible, and interpretable method for obtaining  LLM-based relevance labels. Its design ensures that the model evaluates information retrieval systems consistently while maintaining flexibility across different datasets, search intents, and language models.

\begin{figure}
\centering
\begin{tcolorbox}[colframe=black, colback=white, coltitle=black, width=\columnwidth, boxrule=0.2mm, arc=3mm,  halign=left]

Given a query and a passage, you must provide a score on an integer scale of 0 to 3 with the following meanings:\par
0 = represent that the passage has nothing to do with the query, \par
1 = represents that the passage seems related to the query but does not answer it, \par
2 = represents that the passage has some answer for the query, but the answer may be a bit unclear, or hidden amongst extraneous information and \par
3 = represents that the passage is dedicated to the query and contains the exact answer.\par
\vspace{10pt}
Important Instruction: Assign category 1 if the passage is somewhat related to the topic but not completely, category 2 if passage presents something very important related to the entire topic but also has some extra information and category 3 if the passage only and entirely refers to the topic. If none of the above satisfies give it category 0.\par
\vspace{10pt}
Query: \{query\}\par
Passage: \{passage\}\par
\vspace{10pt}
Split this problem into steps:\par
Consider the underlying intent of the search.\par
Measure how well the content matches a likely intent of the query (M).\par
Measure how trustworthy the passage is (T).\par
Consider the aspects above and the relative importance of each, and decide on a final score (O). Final score must be an integer value only.\par
Do not provide any code in result. Provide each score in the format of: \#\#final score: score without providing any reasoning.
\end{tcolorbox}
\caption{The zero-shot Bing (UMBRELA) prompt for relevance assessment. All prompts used in this experiment are sourced from the original UMBRELA GitHub repository and presented as-is (see main text for details).} 
\label{fig:umbrela_prompt}
\end{figure}

\subsection{Reproduction Overview}

In this work, we reproduce the \textbf{UMBRELA} evaluation framework \cite{upadhyay2024umbrelaumbrelaopensourcereproduction}, originally tested only on GPT-4 family models, and extend it to other large and smaller open-source LLMs. To ensure reproducibility, we:

\begin{itemize}
    \item \textbf{Reuse}: We reuse the zero-shot UMBRELA relevance prompt (Figure~\ref{fig:umbrela_prompt}) form the original paper without modification. We also reuse the zero-shot Basic prompt (Figure~\ref{basic_prompt_figure}) from the UMBRELA repository. All prompts are sourced from the original GitHub repository.\footnote{\url{https://github.com/castorini/umbrela/tree/main/src/umbrela/prompts}}
    \item \textbf{Re-implement}: For the re-implementation of umbrela with other LLMs, we implement our own code to load TREC DL datasets and experiments with different LLMs (both locally via Hugging Face \footnote{https://huggingface.co/} and API-based). We build the full inference and evaluation pipeline.
    \item \textbf{Copy Results}: We copy the \gpto{} results for comparison directly from Table 2 of Upadhyay et al. \cite{upadhyay2024umbrelaumbrelaopensourcereproduction} as the baseline for comparison and keep them unchanged to maintain consistency with the original study.
\end{itemize}

This setup allows us to validate UMBRELA results and test generalization across models of varying capabilities and access levels.

\begin{figure}
\centering
\begin{tcolorbox}[colframe=black, colback=white, coltitle=black, width=\columnwidth, boxrule=0.2mm, arc=3mm,  halign=left]

You are an expert judge of a content. Using your internal knowledge and simple commonsense reasoning, try to verify if the passage is relevance category to the query.
Here, "0" represent that the passage has nothing to do with the query, "1" represents that the passage seems related to the query but does not answer it, "2" represents that the passage has some answer for the query, but the answer may be a bit unclear, or hidden amongst extraneous information and "3" represents that the passage is dedicated to the query and contains the exact answer.\par
\vspace{10pt}

Provide explanation for the relevance and give your answer with from one of the categories 0, 1, 2 or 3 only. One of the categorical values if compulsory in answer.\par
\vspace{10pt}

Instructions: Think about the question. After explaining your reasoning, provide your answer in terms of 0, 1, 2 or 3 category. Only provide the relevance category on the last line. Do not provide any further details on the last line.\par
\vspace{10pt}

\#\#\#\par
\vspace{10pt}

Query: \{query\}\par
Passage: \{passage\}\par
\vspace{10pt}

Explanation:\par
\end{tcolorbox}
\caption{The zero-shot basic prompt for relevance assessment. All prompts used in this experiment are sourced from the UMBRELA GitHub repository and presented as-is (see main text for details).}
\label{basic_prompt_figure}
\end{figure}

\subsection{Task Description}

In line with the original UMBRELA paper, we adopt the same relevance assessment task defined by the TREC Deep Learning (DL) Track. Following the reproducibility goal of our study, we retain this task formulation to ensure comparability with the original results. 

\begin{quote}
The task involves assigning a relevance label to a query-passage pair using a four-point ordinal scale, as defined by human NIST assessors. The scale ranges from [0] (irrelevant) to [3] (perfectly relevant) and was originally created for the TREC Deep Learning track \cite{dl19}.

\begin{itemize}
  \item \textbf{[3] Perfectly Relevant}: The passage directly answers the query with exact information.
  \item \textbf{[2] Highly Relevant}: The passage contains some answer to the query, but the answer may be unclear or embedded in extraneous information.
  \item \textbf{[1] Related}: The passage is somewhat related to the query but does not provide a direct answer.
  \item \textbf{[0] Irrelevant}: The passage has no relevance to the query.
\end{itemize}
\end{quote}
These labels represent the ground truth annotations provided by human assessors and serve as the expected outputs for the relevance prediction task. The evaluation is based on how closely model-predicted labels match these human-annotated judgments.

While this formulation is designed for the TREC DL task, applying our approach to other evaluation settings,such as conversational relevance or multi-turn QA, requires adapting the prompt to reflect the specific criteria and label definitions of those tasks.

\begin{table}
    \centering
    \caption{Dataset Statistics}
    \label{tab:statistics}

\begin{tabular}{ccccccc}
\toprule
 &  &  & \multicolumn{4}{c}{Relevance Label Counts}\tabularnewline
 & Systems & Queries & 0 & 1 & 2 & 3\tabularnewline
 \midrule
\dlnineteen{} & 36 & 43 & 5158 & 1601 & 1804 & 697\tabularnewline
\dltwenty{} & 59 & 54 & 7780 & 1940 & 1020 & 646\tabularnewline

\llmjudge{}  & 35 & 25 & 2005 & 1233 & 808 & 377  \tabularnewline
\bottomrule
\end{tabular}
\end{table}

\subsection{Datasets}

This framework is evaluated on the same year \textbf{TREC Deep Learning datasets} used in the original UMBRELA evaluation \cite{upadhyay2024umbrelaumbrelaopensourcereproduction}.

\begin{description}
    \item[\llmjudge{} (primary dataset):] Systems submitted to TREC DL 2023 with duplicates removed \cite{rahmani_llmjudge_2024}.\footnote{Cleaned \llmjudge{} data available at \url{https://llm4eval.github.io/LLMJudge-benchmark/}}\\
    We primarily focus our analysis on the \llmjudge{} dataset, as it represents the most recent public test collection and minimizes potential data leakage during LLM training \cite{clarke2024llmbasedrelevanceassessmentcant}.
    
    \item[\dltwenty{}:] Systems for passage retrieval from the TREC Deep Learning Track 2020 \cite{dl20}.
    
    \item[\dlnineteen{}:] Systems for passage retrieval from the TREC Deep Learning Track 2019 \cite{dl19}.
\end{description}

Statistics about these datasets, including the number of submitted systems, queries, and the distribution of label statistics, are provided in Table~\ref{tab:statistics}.

This experimental setup allows us to compare our results with those reported in the original paper.

\begin{table*}
\centering
\small
\caption{
Comparison of UMBRELA variants across different LLMs. Results are based on Cohen's $\kappa$ scores (four-scale as well as 01-vs-34 and their correlation with ground-truth judgments.  Corelation metrics Spearman's rank ($\rho$) and Kendall's Tau ($\tau$) are used to compare the leaderboards. Following track guidelines, the evaluation uses nDCG@10 as the primary system evaluation metric.  An asterisk (*) denotes results reported in the original paper. 
% A plus sign (+) indicates datasets where a preprocessing step was applied to remove near-duplicate passages. 
The DL 2023 dataset used in our experiments corresponds to the LLMJudge challenge data \cite{rahmani_llmjudge_2024}. All approaches use Zeroshot Bing prompt (Figure \ref{fig:umbrela_prompt}) unless otherwise noted. 
The bottom provides additional results from the literature for comparison. Best results (or equivalent) are marked in bold.
}
\label{tab:llm_comparison}

% Preview source code for paragraph 0

% Preview source code for paragraph 5
% \begin{small}
\begin{tabular}{lccccccccccccccc}
\toprule 
 & \multicolumn{4}{c}{\dlnineteen{}} &  & \multicolumn{4}{c}{\dltwenty{}} &  & \multicolumn{4}{c}{\llmjudge{}} & \tabularnewline
\midrule 
 & \multicolumn{2}{c}{Rank Correlation} & \multicolumn{2}{c}{Cohen $\kappa$} &  & \multicolumn{2}{c}{Rank Correlation} & \multicolumn{2}{c}{Cohen $\kappa$} &  & \multicolumn{2}{c}{Rank Correlation} & \multicolumn{2}{c}{Cohen $\kappa$} & \tabularnewline
 & $\rho$  & $\tau$  & scale  & binary  &  & $\rho$  & $\tau$  & scale  & binary  &  & $\rho$  & $\tau$  & scale  & binary  & \tabularnewline
\midrule 
\gpto{}{*}  & \textbf{0.974}  & \textbf{0.893}  & \textbf{0.361}  & 0.499  &  & \textbf{0.992}  & \textbf{0.943}  & \textbf{0.351}  & \textbf{0.450}  &  & 0.985  & 0.911  & \textbf{0.308}  & \textbf{0.418}  & \tabularnewline
\deepseek{}  & \textbf{0.978}  & \textbf{0.898}  & \textbf{0.360}  & \textbf{0.518}  &  & \textbf{0.993}  & \textbf{0.940}  & \textbf{0.358}  & 0.426  &  & \textbf{0.989}  & 0.929  & 0.262  & 0.371  & \tabularnewline
\llamabig{}  & 0.970  & 0.869  & 0.266  & 0.447  &  & 0.986  & 0.911  & 0.247  & 0.344  &  & \textbf{ 0.993}  & \textbf{0.946}  & 0.233  & 0.391  & \tabularnewline
\llamasmall{}  & \textbf{0.975}  & \textbf{0.894}  & 0.108  & 0.244  &  & 0.973  & 0.870  & 0.115  & 0.187  &  & \textbf{0.989}  & 0.931  & 0.187  & 0.315  & \tabularnewline
\tfive{}  & 0.964  & 0.862  & 0.050  & 0.292  &  & 0.880  & 0.717  & 0.036  & 0.231  &  & 0.971  & 0.868  & 0.062  & 0.209  & \tabularnewline
\midrule 
%\tabularnewline
%\tfive Thomas &  & &  & &  &  &  & 0.751 & 0.623 &  &  & &  &  & \tabularnewline
%\midrule
% \multicolumn{3}{l}{\tfive{} Faggioli \cite{dietz2024workbench}} &  0.864 & 0.810& &  &  & 0.940 & 0.815 &  &  & &  &  & \tabularnewline
 &  &  &  &  &  &  &  &  &  &  &  &  &  &  & \tabularnewline
\end{tabular}% \end{small}

\end{table*}

\subsection{LLM Selection and Inference Setup}

We select models from different families to study the effects of using the UMBRELA evaluation framework when \gpto{} is replaced.

\begin{description}
    \item[\gpto{} (used in original study):] Proprietary model offered via OpenAI’s API.\footnote{\url{https://platform.openai.com/docs/models}}  
     The architectural leak reported by SemiAnalysis estimates GPT-4 as a MoE model with 16 experts, 2 active per token, each $\approx 111B$ parameters.
     We include it as the reference point for comparison because it was used in the original study, and based on this reproduction, we would advise users to use this model with UMBRELA.
    \item[\deepseek{}:] Recently released open-source model with reasoning enhancements distilled from DeepSeek-R1 \cite{DeepSeekAI2024DeepSeekV3TR}, known to be trained with data generated by \gpto{}, making it a strong publicly available contender.\footnote{\url{https://www.together.ai/models/deepseek-v3}} The total model size is $\approx 145B$, with MoE architecture implied; active parameter usage per token is about 2.8B.\footnote{\url{https://milvus.io/ai-quick-reference/what-is-the-parameter-count-of-deepseeks-r1-model}}
    \item[\llamabig{}:] Large-scale open-source model, included to test performance at higher capacity.\footnote{\url{https://api.together.xyz/models/meta-llama/Llama-3.3-70B-Instruct-Turbo}}  A dense transformer model with 70B parameters, no experts.
    This model is developed independently from \gpto{}, ensuring distinct training processes.
    \item[\llamasmall{}:] Open-source model with medium-scale deployment constraints (8B), chosen for its instruction-tuning and accessibility for resource-constrained settings and because it is often compared to \llamabig{}  \cite{takehi2024llm}.\footnote{\url{https://huggingface.co/meta-llama/Meta-Llama-3-8B-Instruct}} 
    \item[\tfive{}:] Open-source, small-scale LLM with 783M parameters. We included it to test the limits of how lightweight a model can be while still supporting reliable evaluation under the UMBRELA framework.\footnote{\url{https://huggingface.co/google/flan-t5-large}}
\end{description}

All models were prompted using the same templates to ensure consistent evaluation.

The original UMBRELA evaluation used OpenAI’s latest \gpto{} model, accessed via Microsoft Azure, to assign relevance labels. For reference, we use the results from \gpto{} as provided in the original UMBRELA paper.

For smaller models, such as \llamasmall{} and \tfive{}, we run experiments locally on an NVIDIA A40 GPU. For larger models, including \deepseek{} and \llamabig{}, we use the Together API\footnote{https://www.together.ai/} for inference. For all models, during inference, we use deterministic decoding by disabling sampling (temperature = 0), ensuring consistency across runs, similar to the original study. Each input was processed individually (batch size = 1) for simplicity and consistency.

\subsection{Evaluation Metrics}

We use the following evaluation metrics to assess model performance, including both leaderboard rank correlation metrics and per-label agreement.

Evaluation measures for comparing the leaderboards resulting from the NDCG@10 evaluation measure under manual judgments and each of the UMBRELA variants:
\begin{description}
    \item[Spearman’s rank correlation ($\rho$):] Measures the difference in systems' rank across the two leaderboards, taking into account each system’s exact rank position.
    \item[Kendall’s tau ($\tau$):] Assesses the consistency of pairwise system orderings between the two leaderboards \cite{Kendall1938ANM}.
\end{description}

Evaluation measures based on per-label agreement:
\begin{description}
    \item[Cohen’s $\kappa$, 4-point scale (scale):] Measures the number of times human judges and UMBRELA agree on the exact label (0--3) for all passages in the judgment pool, corrected for chance agreement.
    \item[Binarized Cohen’s $\kappa$ (binary):] The same measure, but with the graded relevance labels binarized: labels 0 and 1 are considered non-relevant, while labels 2 and 3 are considered relevant.
\end{description}

\paragraph{Significance} While significance tests do not apply to these measures, we assume that  a difference of $\leq 0.005$ for $\tau$ and $\rho$ and $\leq 0.01$ for Cohen’s $\kappa$ is not significant. The best values and equivalently good results are highlighted in bold.

\bigskip
\bigskip
\section{Results}

\subsection{Comparison of Different LLMs in the UMBRELA Framework}

First, we address the most pressing question: can we replace the \gpto{} LLM in the UMBRELA framework and still expect equally good results? This is formalized in the following research question:

\begin{quote}
    \textbf{RQ1: Can we replace the LLM in UMBRELA without performance degradation?}  
    --- Finding: Technically yes, but its reliability varies with model scale.
\end{quote}

Different LLMs using the original UMBRELA prompt (Figure~\ref{fig:umbrela_prompt}) are compared in Table~\ref{tab:llm_comparison}. We find that the \gpto{} model usually performs much better than other LLMs, with \deepseek{} showing very similar performance, as the second-best performer. In \llmjudge{}, \deepseek{} performs slightly worse than \gpto{} in Cohen's $\kappa$ but outperforms it in $\tau$ and $\rho$.

The overall third-best system is UMBRELA with either \llamabig{} or \llamasmall{}, though their per-label agreement metrics ($\kappa$) are consistently lower. However, the impact on the leaderboard ranking is small, and leaderboard rankings of both \llama{} LLMs closely resemble the original leaderboard based on human judgments.  

We note that the smallest (and fastest) LLM, \tfive{}, despite lower per-label agreement, still obtains strong performance in terms of leaderboard rank correlation.

\bigskip
\bigskip

\subsection{Leaderboard vs.\ Per-label Agreement}
Per-label agreement measures and leaderboard rank correlation measures reflect different evaluation goals.

\begin{quote}
    \textbf{RQ2: To what extent do leaderboard rank correlation and per-label agreement metrics differ in their sensitivity to LLM scale?}  
    --- Finding: While all LLMs produce leaderboards closely resembling those of human judges, differences are more noticeable in per-label measures.
\end{quote}

\begin{figure}
    \centering
    \includegraphics[width=1\linewidth]{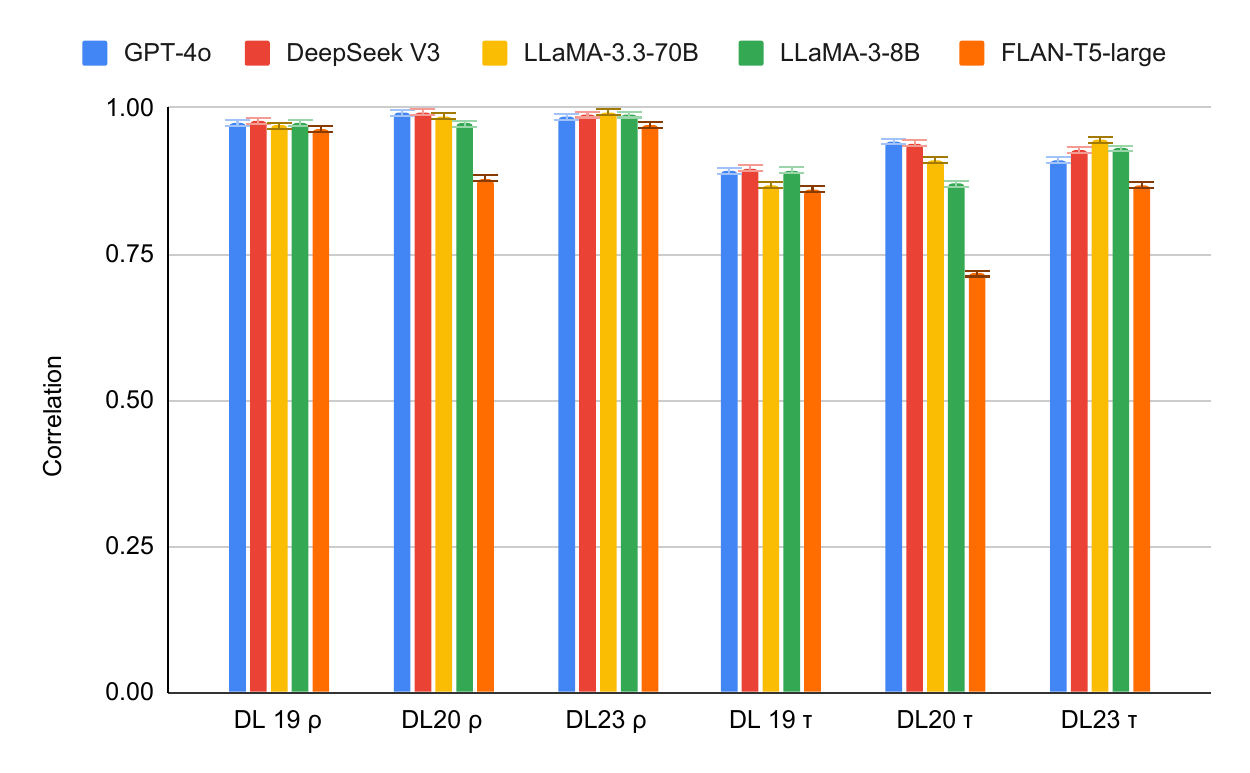}
    \caption{Effect of model scale on rank correlation metrics.}
    \label{fig:plot-scale-effect-rank-correlation}
\end{figure}

In general, we find that all evaluation measures are in agreement on the best LLM-Judge systems (\deepseek{} and \gpto{}), runner-up (\llamabig{}), and weaker systems (\llamasmall{} and \tfive{}), as depicted in Figures~\ref{fig:plot-scale-effect-rank-correlation} and~\ref{fig:plot-scale-effect-interannotator-agreement} and Table~\ref{tab:llm_comparison}.

\paragraph{UMBRELA on \gpto{}.}
On the \llmjudge{} collection, UMBRELA with \gpto{} achieves a Cohen’s $\kappa$ of 0.31 on the four-point relevance scale. This aligns with results from the LLM Judge Challenge at SIGIR 2024 \cite{rahmani_llmjudge_2024}, where UMBRELA variants reported scale-level $\kappa$ scores of 0.29 and 0.28. Similarly, our observed binary $\kappa$ of 0.42 is consistent with the 0.40 and 0.43 obtained by UMBRELA variants at the LLM Judge challenge.

\paragraph{Scale Sensitivity of Per-Label Agreement.}  
Looking at how the UMBRELA prompt performs across models of different scales in Table~\ref{tab:llm_comparison}, we observe distinct patterns in how per-label agreement (Cohen's $\kappa$) and rank correlation metrics (Spearman's $\rho$, Kendall's $\tau$) respond to changes in model size. The per-label agreement metrics benefit more substantially from increased model scale than the leaderboard rank correlation measures. This is particularly evident when comparing the largest models (\gpto{} and \deepseek{}) to smaller ones like \tfive{}.  

For instance, in \llmjudge{}, the UMBRELA prompt on \gpto{} obtains a four-point scale Cohen’s $\kappa$ of 0.308, which drops dramatically to 0.062 for \tfive{}, representing an 80\% decrease. In contrast, the rank correlation metrics are remarkably robust: $\rho$ decreases only from 0.985 to 0.971 (a 1.4\% drop), and $\tau$ from 0.911 to 0.868 (a 4.7\% drop). This pattern is consistent across all three datasets, \dlnineteen{}, \dltwenty{}, and \llmjudge{}.

\paragraph{Binary vs.\ Scale $\kappa$.}  
Among the per-label measures, the binary version of Cohen’s $\kappa$ consistently yields higher values due to a greater likelihood of chance agreement. However, this gap varies with model scale---larger models like \gpto{} and \deepseek{} show smaller differences between their binary and scale $\kappa$ scores compared to smaller models. This indicates that increased model scale particularly benefits fine-grained judgment capabilities, especially in distinguishing between non-relevant labels (0 and 1) and relevant labels (2 and 3).

\bigskip
These findings suggest that when accurate per-label judgments are required---such as for generating synthetic training sets or fine-grained evaluations---UMBRELA should only be used with larger models like \gpto{} and \deepseek{}.

\begin{table*}
\centering

% Preview source code for paragraph 5
% \begin{small}

\centering

\caption{
Comparison of the suggested UMBRELA prompts to the alternative ``Basic Prompt'' using the \llamasmall{} model. Same information as in Table \ref{tab:llm_comparison}. 
}
\label{tab:prompt_comparison}

\begin{tabular}{lccccccccccccccc}
\toprule
 & \multicolumn{4}{c}{\dlnineteen{}} &  & \multicolumn{4}{c}{\dltwenty{}} &  & \multicolumn{4}{c}{\llmjudge{} +} & \tabularnewline
\cline{2-5}   \cline{7-10} \cline{12-15}  
 & \multicolumn{2}{c}{Cohen $\kappa$} & \multicolumn{2}{c}{Correlation} &  & \multicolumn{2}{c}{Cohen $\kappa$} & \multicolumn{2}{c}{Correlation} &  & \multicolumn{2}{c}{Cohen $\kappa$} & \multicolumn{2}{c}{Correlation} & \tabularnewline
 & scale & binary & $\rho$ & $\tau$ &  & scale & binary & $\rho$ & $\tau$ &  & scale & binary & $\rho$ & $\tau$ & \tabularnewline
 \midrule

\llamasmall{} (Basic) & \textbf{0.108} & 0.233 & 0.934 & 0.786 &  & \textbf{0.126} & \textbf{0.183} & \textbf{0.981} & \textbf{0.901} &  & 0.173 & 0.250 & \textbf{0.988} & \textbf{0.932} & \tabularnewline
\llamasmall{} & \textbf{0.108} & \textbf{0.244} & \textbf{0.975} & \textbf{0.894} &  & 0.115 & \textbf{0.187} & 0.973 & 0.870 &  & \textbf{0.187} & \textbf{0.315} & \textbf{0.989} & \textbf{0.931} & \tabularnewline

\bottomrule
\end{tabular}
% \end{small}

\end{table*}

\bigskip

\subsection{Generalization Across Datasets}
\label{sec:result-datasets}
Next, we examine whether the findings generalize across datasets.

\begin{quote}
    \textbf{RQ3: How consistent are UMBRELA’s performance patterns across different relevance assessment datasets, and does this consistency vary with model scale?}  
    --- Finding: Larger models like \gpto{} and \deepseek{} achieve the best results across all datasets.
\end{quote}

Examining Table~\ref{tab:llm_comparison}, we observe that UMBRELA’s performance exhibits meaningful patterns of consistency across the three datasets \dlnineteen{}, \dltwenty{}, and \llmjudge{}, though with some notable variations.

For larger models like \gpto{} and \deepseek{}, performance remains relatively stable across datasets. \gpto{} maintains strong binary Cohen’s $\kappa$ scores (0.499, 0.450, 0.418) and high rank correlation metrics ($\rho$ ranging from 0.974 to 0.992, and $\tau$ ranging from 0.893 to 0.943) across all test collections. \deepseek{} performs very similarly,  obtaining binary $\kappa$ scores of 0.518, 0.426, and 0.371 with comparable $\rho$ ranging from 0.978 to 0.993 and $\tau$ between 0.898 and 0.940.

\llamabig{} usually places third, with the exception of \llmjudge{}, where it outperforms \deepseek{} in terms of binary $\kappa$ and \gpto{} in terms of Spearman's $\rho$ and Kendall's $\tau$. Also, we note \llamasmall{} slightly outperforms \llamabig{} in terms of rank correlation for \dlnineteen{}.

While \tfive{} (the smallest model) usually places last, we find that it outperforms \llamasmall{} in terms of binary $\kappa$ for \dlnineteen{} and \dltwenty{}.

Interestingly, in terms of rank correlation, models across all scales perform remarkably well, resulting in nearly indistinguishable performance.  Even for the small \tfive{},  we find that Spearman's rank correlation coefficient $\rho$ on the \llmjudge{} and \dlnineteen{} datasets is nearly indistinguishable from that of \gpto{}.
The exception is \tfive{}'s reduced performance on \dltwenty{} resulting in $\rho=0.880$ and $\tau=0.717$.  Notably,  shorter prompts \citep{Faggioli2023PerspectivesOL,Thomas2023LargeLM,Sun2023IsCG} perform better for this LLM, achieving $\rho$ of 0.94--0.97
and $\tau$ of 0.81--0.88
according to Farzi and Dietz~\citep{farzi2024pencils}.

Future research should explore whether the lower performance on small-scale models \llamasmall{} and \tfive{} is inherent to their limited capacity, or whether the UMBRELA prompt is simply not well-suited to smaller LLMs.

\bigskip
\bigskip

\subsection{Are Larger LLMs Better?}
\label{RQ4}
We assess the extent to which UMBRELA benefits from large-scale LLMs.

\begin{quote}
    \textbf{RQ4: To what extent does investing in larger language models provide meaningful improvements in relevance assessment quality beyond leaderboard rank correlation?}  
    --- Finding: UMBRELA performs better with larger LLMs, especially when per-label agreement is important.
\end{quote}

\begin{figure}
    \centering
    \includegraphics[width=1\linewidth]{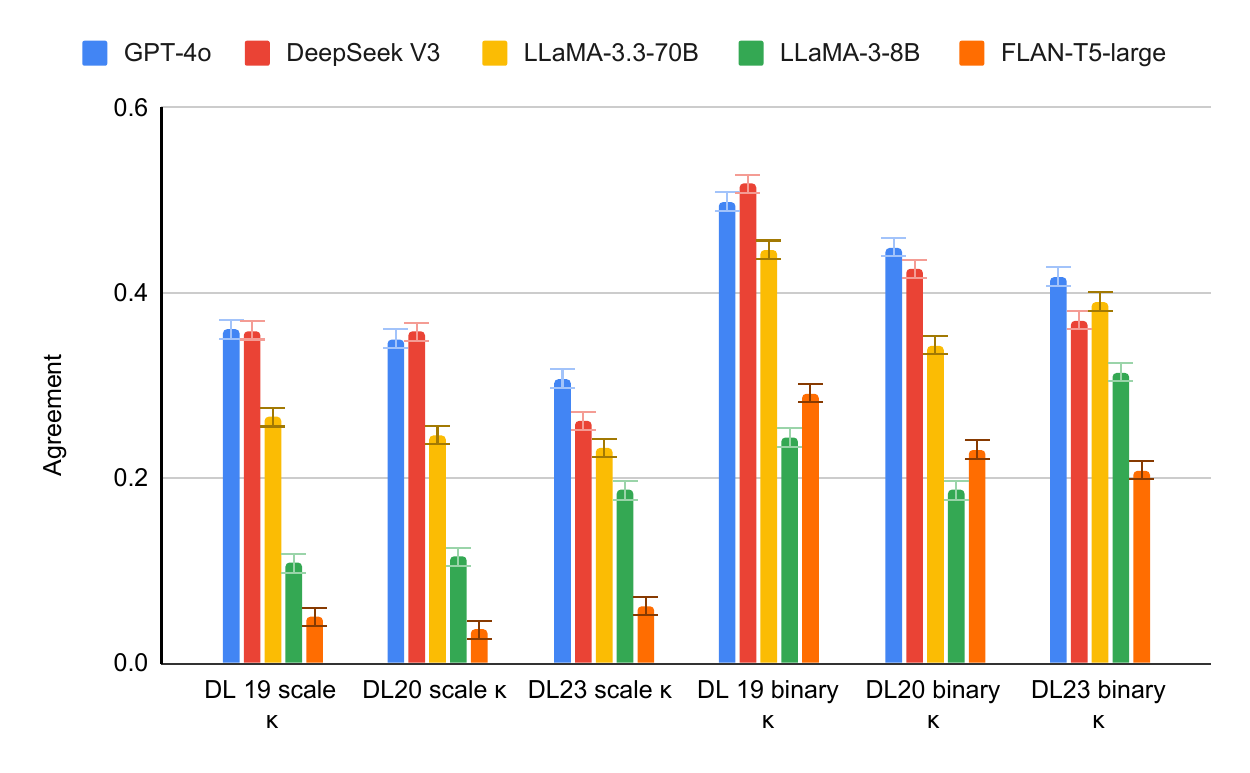}
    \caption{Effect of model scale on per-label agreement.}
    \label{fig:plot-scale-effect-interannotator-agreement}
\end{figure}

Across the board, UMBRELA on any model achieves strong results above 0.85 in terms of rank correlation metrics ($\rho$ and $\tau$). This is encouraging, as it implies that all model variants can differentiate between top-performing and lower-performing systems. This finding even holds for UMBRELA with the small-scale \tfive{} (except on \dltwenty{}). Hence, for some evaluation tasks, smaller and more efficient models may suffice.

However, if the goal is to derive a synthetic training set, then per-label agreement measures become more important. In this case, \gpto{} and \deepseek{} should be preferred,  as depicted in  Figure~\ref{fig:plot-scale-effect-interannotator-agreement}.

\begin{table}
\centering
\caption{Invalid output percentages for models across datasets using UMBRELA prompt}
\begin{tabular}{lc}
\toprule
Model & Invalid Outputs (\%) \\
\midrule
\deepseek{} & 0.00 \\
\llamabig{} & 0.00 \\
\llamasmall{} & 1.59--1.87 \\
\tfive{} & 0.00--0.05 \\
\bottomrule
\end{tabular}
\label{tab:invalid-outputs}
\end{table}

\subsection{Choice of Prompt}
Much research has focused on tuning the LLM prompt, with less attention given to designing prompts tailored to specific LLM families.

\begin{quote}
    \textbf{RQ5: How much does the word choice of the prompt matter in comparison to the LLM?}  
    --- Finding: We find that slight differences in prompt wording do not lead to noticeable quality changes.
\end{quote}

 The UMBRELA setup also includes an alternative ``Basic Prompt,'' shown in Figure~\ref{basic_prompt_figure}, which is available in the UMBRELA GitHub repository\footnote{\url{https://github.com/castorini/umbrela/tree/main/src/umbrela/prompts}} but not discussed in the original paper. The ``Basic Prompt'' omits instructions on query intent and trustworthiness, and hence serves as a simpler baseline. We compare prompt performance on \llamasmall{}, the medium-scale LLM that balances cost and performance.
 Results in Table~\ref{tab:prompt_comparison} reveal dataset-specific differences for these two prompts.
 We find that both prompts yield comparable performance overall: the UMBRELA prompt performs slightly better on \dlnineteen{} and \llmjudge{}, while the ``Basic Prompt'' performs better on \dltwenty{}. This corroborates our suspicion that shorter prompts may work better with smaller LLMs (cf.\ discussion in Section~\ref{sec:result-datasets}). 

We conclude that while prompt wording can yield small differences, the choice of LLM has a greater impact on UMBRELA's performance.

\subsection{Output Errors and Handling}
It is easy to dismiss the lower meta-evaluation scores as merely the results if a less capable model. However, based on our error analysis, we found that a significant contributing factor to performance differences was models' ability to follow the prescribed output format. Despite explicit instructions to provide relevance judgments in the format ``\texttt{\#\#final score: [score]}'', smaller models like \tfive{} only generated standalone numbers as the relevance label, while medium-scale models like \llamasmall{} frequently produced inconsistent outputs---ranging from component scores without mentioning the word ``Final Score'' (e.g., \texttt{M: 3, T: 3, O: 3}) to unformatted numbers. Larger models like \llamabig{} occasionally deviated as well, generating step-by-step reasoning or alternative formatting, which necessitated more complex parsing logic and introduced extraction errors.

To extract the final UMBRELA relevance score (0--3) from LLM outputs, we use a simple, rule-based pattern-matching function that checks for specific score formats, including \texttt{``\#\#final score: X''}, \texttt{``O: X''}, and a fallback standalone number. The prompt explicitly instructs the model to use the \texttt{``\#\#final score: X''} format, and we also accommodate the \texttt{``O: X''} format as a secondary fallback. This is based on the prompt's instruction where the final score is referred to as ``O'', and some models may adopt this shorthand to label the relevance score. Additionally, the function can handles minor formatting variations, such as spacing between \texttt{``\#\#``} and \texttt{``final score``}. If none of the expected patterns are found, a default score of 0 is returned.

Using the UMBRELA prompt, we find 0.00\% erroneous output for \deepseek{} and \llamabig{}, but approximately 1.7\% for \llamasmall{}, and up to 0.05\% for \tfive{} across datasets (see Table~\ref{tab:invalid-outputs}). See Appendix~\ref{sec:output-errors} for sample LLM outputs. LLMs that are unreliable in instruction following have the potential to negatively impact LLM-based evaluations.

\subsection{Worked Example of Different LLMs}
\label{app:same-query-diff-llms}

\begin{table}[h!]
\centering
\caption{Outputs of different LLMs on the same query and passage pair (qidx=555530, docidx=58950, Relevance Label=2).}
\label{tab:llm-comparison}
\begin{tabular}{@{}lp{5.5cm}@{}}
\toprule

\textbf{Input} & \textbf{Text} \\ \midrule
Query & what are best foods to lower cholesterol \\ \midrule
Passage & There are two forms of fiber: soluble and insoluble. Soluble fiber attracts water and turns to gel during digestion. This slows digestion. Soluble fiber is found in oat bran, barley, nuts, seeds, beans, lentils, peas, and some fruits and vegetables. Research has shown that soluble fiber lowers cholesterol, which can help prevent heart disease. Insoluble fiber is found in foods such as wheat bran, vegetables, and whole grains. \\ \midrule
\textbf{Model} & \textbf{Output} \\ \midrule

\llamasmall{} & Here are the scores:
\\
&M: 2\\
&T: 3\\
&O: 2 \\ \midrule
\llamabig{} & \#\#M: 2\\
&\#\#T: 2\\
&\#\#O: 2 \\ \midrule
\deepseek{} & \#\#final score: 2 \\ \midrule
\tfive{} & 3 \\ \bottomrule
\end{tabular}
\end{table}

We demonstrate how relevance labels are predicted by the UMBRELA prompt with different language models in Table~\ref{tab:llm-comparison}. We compare the outputs of four LLMs---\llamasmall{}, \llamabig{}, \deepseek{}, and \tfive{}---on query \texttt{555530} and passage \texttt{58950} from the \dltwenty{} dataset. According to manually created relevance labels, this passage is ``Highly Relevant'', corresponding to label code $2$, although one could argue that a higher rating might also be justified.

We observe that the models’ output formats, and where applicable, their decomposition into sub-scores, vary. Nevertheless, all LLMs are able to identify this passage as relevant for the query. In this case, most LLMs are correctly predicting the relevance label as $2$, while \tfive{} predicts it as ``Perfectly Relevant'' ($3$).

\section{Conclusion}
Our study demonstrates that UMBRELA can operate across different LLMs, though with notable performance variations. \gpto{} and \deepseek{}  achieve the highest per-label agreement metrics. However, even smaller models (e.g., \tfive{}) produce leaderboards that closely resemble the official rankings. This highlights a key distinction: with the current prompt, leaderboard correlations remain stable across model scales, while per-label agreement metrics are highly sensitive to model scale.

Across all datasets, we find that the large-scale LLMs work better with the UMBRELA prompt. However, we also observe diminishing returns in cost-benefit tradeoffs as model scale increases, suggesting that medium-sized models, such as both \llama{} versions, may offer an optimal balance between performance and efficiency for many applications.

These findings suggest that UMBRELA implementation choices should be guided by specific evaluation goals: for system ranking, smaller models may suffice, while accurate document-level relevance labels require larger models. This nuanced understanding of LLM-based evaluation provides practical guidance for IR researchers seeking to balance computational efficiency with evaluation integrity.

\begin{acks}
We thank the authors of the original UMBRELA framework \cite{upadhyay2024umbrelaumbrelaopensourcereproduction}  for providing access to prompts and baseline results. We are also grateful to the organizers of the TREC Deep Learning Track \cite{dl19,dl20} for making the test collections available, and to the organizers of the LLM Judge Challenge \cite{rahmani_llmjudge_2024} for providing a valuable testbed for experimentation. 

This material is based upon work supported by the National Science Foundation under Grant No. 1846017. Any opinions, findings, and conclusions or recommendations expressed in this material are those of the author(s) and do not necessarily reflect the views of the National Science Foundation.
\end{acks}

\appendix
\section{Appendix: Output Examples}
\label{sec:output-errors}

In this appendix, we provide examples of output formats from various LLMs where the requested output format was not consistently followed. Despite using the same LLM and prompt, some models produced inconsistent outputs. The examples presented here illustrate some types of inconsistencies observed.

\subsection{\llamasmall{}}
The following examples highlight different output formats generated by \llamasmall{}:

\begin{itemize}
    \item Example 1: \\ \texttt{M: 3\\
    T: 3\\
    O: 3}
    \item Example 2: \\ \texttt{Here are the scores:\\
    M: 3\\
    T: 3\\
    O: 3}
    \item Example 3: \\
    \texttt{ 1: 2}
    \item Example 4: \\ 
    \texttt{0}
\end{itemize}

\subsection{\llamabig{}}
Examples of inconsistent output formats from \llamabig{} include:

\begin{itemize}
    \item Example 1: \\
    \texttt{ \#\# Step 1: Measure how well the content matches a likely intent of the query (M): 3\\
    \#\# Step 2: Measure how trustworthy the passage is (T): 3\\
    \#\# Step 3: Final score (O): 3}
    \item Example 2: \\ \texttt{ \#\# Step 1: Consider the underlying intent of the search\\
    The underlying intent of the search is to understand how a bounty hunter generates income.\\
    \#\# Step 2: Measure how well the content matches a likely intent of the query (M)\\
    M: 2\\
    \#\# Step 3: Measure how trustworthy the passage is (T)\\
    T: 2\\
    \#\# Step 4: Consider the aspects above and the relative importance of each, and decide on a final score (O)
    O: 2 \\
    \#\#final score: 2}
    \item Example 3:\\
    \texttt{ \#\# M: 3\\
    \#\# T: 2 \\
    \#\# O: 3 \\
    \#\# final score: 3}
    \item Example 4:\\
    \texttt{\#\# Step 1: M = 3\\
    \#\# Step 2: T = 3\\
    \#\# Step 3: O = 3\\
    \#\# final score: 3}
\end{itemize}

\subsection{\deepseek{}}
Although \deepseek{} followed the desired format most of the time (``\#\#final score: X''), it also exhibited other output formats, such as:

\begin{itemize}
    \item Example 1: \\
    \texttt{\#\#M: 1  \\
    \#\#T: 1  \\
    \#\#O: 1}
    \item Example 2: \\
    \texttt{
    \#\#M: 0  \\
    \#\#T: 1  \\
    \#\#O: 0  \\
    \#\#final score: 0
    }
\end{itemize}

\balance

These examples demonstrate that even with identical prompts and models, the output format can vary significantly, highlighting a challenge in ensuring consistency when using LLMs for evaluation tasks.

\bibliographystyle{plain}
\bibliography{sample}
\balance

\end{document}